%% file: main.tex
\documentclass[10pt,twocolumn,letterpaper]{article}

\usepackage[pagenumbers]{cvpr} 
\usepackage{pifont}
\usepackage{graphicx}
\usepackage{caption}
\usepackage{subcaption}
\usepackage{algorithm}
\usepackage{array}
\usepackage[noend]{algpseudocode}
\usepackage[table,x11names]{xcolor}
\usepackage{multirow}

\definecolor{cvprblue}{rgb}{0.21,0.49,0.74}
\usepackage[pagebackref,breaklinks,colorlinks,citecolor=cvprblue]{hyperref}

\title{FSSUAVL: A Discriminative Framework using Vision Models for Federated Self-Supervised Audio and Image Understanding}

\author{Yasar Abbas Ur Rehman, 
\space Kin Wai Lau,
\space Yuyang Xie,  
\space Ma Lan,
\space Jiajun Shen\\
\normalsize TCL AI Lab, Hong Kong
}

\newcommand{\xmark}{\ding{55}}

\begin{document}
\maketitle
\input{sec/0_abstract}    
\input{sec/1_intro}
\input{sec/2_related_work}

\input{sec/3_method}

\input{sec/4_experiment}

\input{sec/6_conclusion}

\clearpage

{
    \small
    \bibliographystyle{ieeenat_fullname}
    \bibliography{bibliography}
}

\end{document}

%% file: sec/0_abstract.tex
\begin{abstract} 
Recent studies have demonstrated that vision models can effectively learn multimodal audio-image representations when paired. 
However, the challenge of enabling deep models to learn representations from unpaired modalities remains unresolved. This issue is especially pertinent in scenarios like Federated Learning (FL), where data is often decentralized, heterogeneous, and lacks a reliable guarantee of paired data.  Previous attempts tackled this issue through the use of auxiliary pretrained encoders or generative models on local clients, which invariably raise computational cost with increasing number modalities. 
Unlike these approaches, in this paper, we aim to address the task of unpaired audio and image recognition using FSSUAVL, a single deep model pretrained in FL with self-supervised contrastive learning (SSL). Instead of aligning the audio and image modalities, FSSUVAL jointly discriminates them by projecting them into a common embedding space using contrastive SSL. This extends the utility of FSSUAVL to paired and unpaired audio and image recognition tasks.
Our experiments with CNN and ViT demonstrate that FSSUAVL significantly improves performance across various image and audio-based downstream tasks compared to using separate deep models for each modality. Additionally, FSSUAVL’s capacity to learn multimodal feature representations allows for integrating auxiliary information, if available, to enhance recognition accuracy. 

\end{abstract}

%% file: sec/1_intro.tex
\section{Introduction}

Federated Self-Supervised Learning (FSSL) has emerged as a promising training paradigm that enables edge devices to collaboratively train a global model from their unlabeled data under the orchestration of a central server keeping their local data private \cite{mcmahan2017communication, kairouz2021advances}. It has been successful in learning joint visual and audio representations from multimodal audio-visual data \cite{feng2023fedmultimodal, feng2025modalitymirror}. However, current practices in FSSL for joint audio and image representation learning are limited to modality alignment, requiring the availability of aligned audio and visual models or a generative model to extract missing modality from available ones (using audio to extract image or vice versa) \cite{feng2023fedmultimodal}. 
Furthermore, unlike centralized settings, where curating a cohesive multimodal audio-visual dataset is feasible, achieving similar alignment across the distributed clients in FSSL is markedly more complex due to inaccessibility to the client's private data. This may give rise to several client-side scenarios:

\noindent\textbf{Independent Datasets:} A client possesses photos and audio tracks lacking any inherent correlation, such as unrelated images and sound clips.

\noindent\textbf{Modality Misalignment:} The audio captures a distinct event from the image, e.g., sound events recorded by a robot’s microphone (behind the robot) that fall outside the visual field of its camera.

\noindent\textbf{Conceptual Mismatch:} The image depicts a forest fire, while the audio features a cheerful pop song, highlighting semantic discordance.

\noindent\textbf{Missing Correspondence:} A dataset includes thousands of animal images but only a few with associated animal sounds, leaving most images without matching audio pairs.

Given these challenging scenarios, FSSL necessitates independent processing of image and audio data to derive robust and meaningful feature representations, often requiring distinct expert models (encoders) for each modality \cite{sun2024towards}. A drawback of training separate expert models for audio and image data is the increased computational and memory demands on the client side.  
Alternatively, one can leverage the techniques of recent works \cite{girdhar2022omnivore} to repurpose a single encoder, such as  CNN or ViT, to simultaneously learn image and audio features using sequential training.  This approach offers flexibility by being both model and modality-agnostic, facilitating efficient scaling across edge devices in FSSL systems. 
Despite these advantages, a single model for audio and image data has rarely been studied in FSSL, and their performance compared to modality-specific models has been disappointing. One of the reasons for such limited performance is that client data may contain an uneven distribution of audio and image data \cite{feng2023fedmultimodal}. Hence, models trained on such data may be tilted toward learning more image-based or audio-based features. This ultimately leads to divergence among the clients' models when aggregated at the server.

To mitigate this issue, we noted that several studies have been conducted to repurpose pretrained vision models, such as CNN \cite{hershey2017cnn, he2016deep} and ViT \cite{gong2022ssast, dosovitskiy2020image}, for audio understanding, leading to the conjecture that these models can be collectively trained on both image and audio data, and that the presence of image data will reinforce improved performance on audio data \cite{gong2022ssast, hershey2017cnn}.  However, these models have never been collectively evaluated on both image and audio data in FSSL. To fill this gap, we propose repurposing a single model for the joint learning of the audio and image modality in FSSL. Our proposed model, termed FSSUVAL, learns global feature representations from the client's unpaired audio and visual data by discriminating them using contrastive SSL and sequential training, similar to the work in \cite{girdhar2022omnivore} (see. Figure \ref{fig:FSSUVAL}). Such a technique projects the feature representations of image and audio into a common embedding space where they are clustered separately. The advantage of such a technique is that the model can work together for both individual and paired modalities when available.    

Unlike prior works \cite{girdhar2022omnivore, girdhar2023imagebind}, we benchmark FSSUVAL by training it on CNNs \cite{he2016deep} and ViT \cite{dosoViTskiy2021an} in FSSL (where SSL is scaled across hundreds of clients). This approach allows for a comprehensive investigation of our approach by scaling it across distributed clients with Non Independently Identically Distributed (non-IID) audio and visual data. It is pertinent to mention here that our work focuses on a more general \textit{cross-device} FL (\textit{where each training iteration trains a random fraction of clients from the pool of clients}) \cite{mcmahan2017communication, rehman2023dawa},  rather than \textit{personalized} FL \textit{(that maintains a global model for each client} \cite{chen2024fedmbridge}) and \textit{cross-silo} FL \textit{(where all the clients train simultaneously} \cite{zhuang2021collaborative}).  

We show that models trained with FSSUVAL can achieve on-par or better performance on both unimodal and multimodal audio and image downstream tasks and require no modification or special treatment for the model architecture. Our experimental results show intriguing observations. For instance, we found that the ViT model pretrained with FSSUVAL without further fine-tuning provides much better performance on audio-based downstream tasks than the modality-specific expert models. When further adapted to modality-specific downstream tasks through transfer-learning and full-network fine-tuning, FSSUVAL, with CNN and ViT, obtains on-par or better performance on both image-based and audio-based downstream tasks. 

%% file: sec/2_related_work.tex
\section{Related Work}
\label{sec:related_work}
FL is characterized by collaborative learning from distributed clients without sharing data, thus respecting the privacy of the client's data \cite{mcmahan2017communication}. In recent years, it has emerged as an alternative method to train machine learning models in a distributed setup due to privacy concerns and the difficulty of keeping data on cloud servers \cite{sani2024future, villalobos2022will}. However, the heterogeneity in the data and computational resources among the clients makes it difficult to achieve performance levels comparable to those achieved in centralized training environments. Despite these shortcomings, recent studies have shown that models trained by combining SSL and FL (FSSL) can surpass the performance of models trained in centralized setups in certain tasks \cite{rehman2022federated, zhuang2021collaborative}. Furthermore, such approaches can also minimize the impact of data heterogeneity on model convergence \cite{zhuang2021collaborative, zhuangdivergence, gao2022federated, lubana2022orchestra}. For example, FSSL has been shown to learn feature representations that are immune to data being IID and Non-IID \cite{rehman2022federated, zhuang2021collaborative}. Although this property has been useful, it has been tested under homogeneous modalities, i.e., all audio or all images. As a result, it is pertinent to investigate whether this property of FSSL holds when each client contains different modalities, i.e., all images, all audio, or a different number of image and audio modalities. 

Previous attempts in multimodal FL aimed to improve a single modality (e.g., image recognition) by using the other modality as complementary information \cite{sun2024towards, zhao2022multimodal}. Furthermore, these works either considered CNN or ViT with paired or unimodal datasets \cite{feng2023fedmultimodal}, where the matching component (text description or image) of the unimodal dataset is found using the additional auxiliary model \cite{chen2024fedmbridge}. Other works \cite{peng2024fedmm} transmit embeddings along with the feature encoders to the server. However, transmitting embeddings might pose privacy risks. Different from these methods, we aim to learn the feature representations from unpaired audio-visual data that can provide better performance on both audio-based and image-based downstream tasks while simultaneously being simple and computationally efficient.

%% file: sec/3_method.tex
\section{Methodology}
\begin{figure*}[htb]
    \centering
    \begin{subfigure}{0.5\linewidth}
    \includegraphics[width=0.9\linewidth]{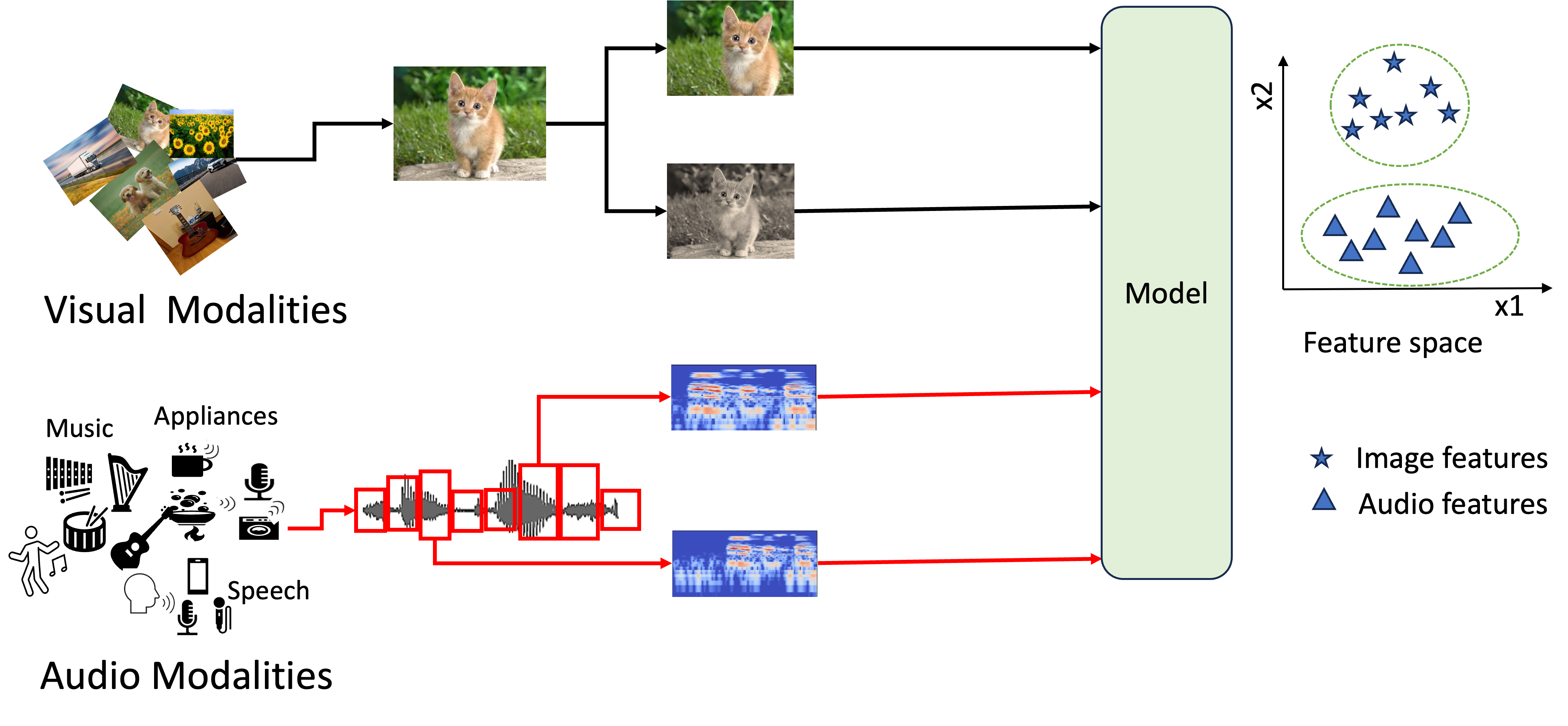}
    \caption{}
    \end{subfigure}%
    \begin{subfigure}{0.5\linewidth}
          \includegraphics[width=0.9\linewidth]{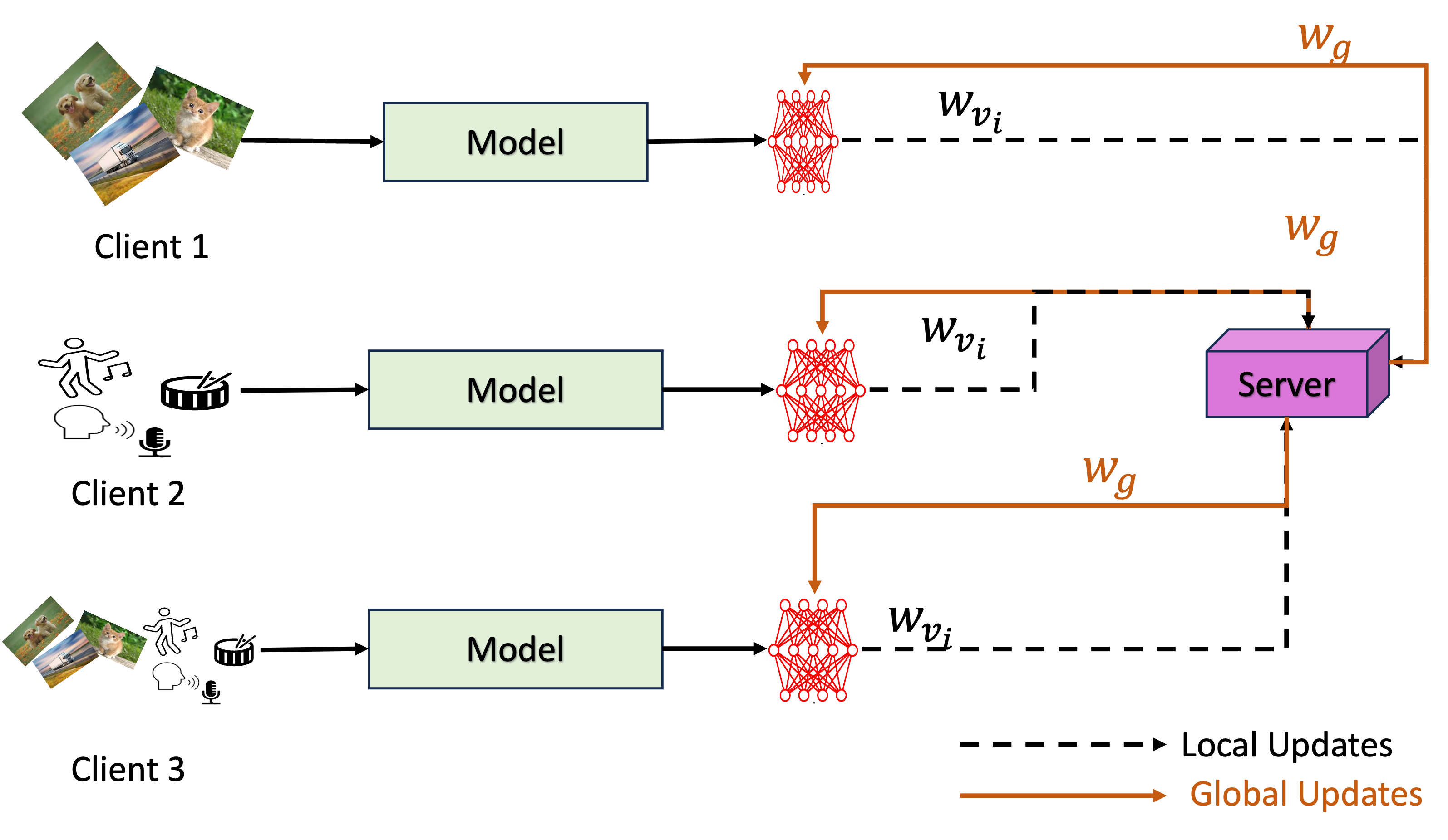}
    \caption{}
    \end{subfigure}
    \caption{FSSUAVL: (a) Local Training. Both modalities are sequentially passed through the same network for representation learning using SSL, which projects them to the common feature space. (b Scaling the local training across multiple clients containing visual-only (client 1), audio-only (client 2), and audio-visual (client 3) data. }
   \label{fig:FSSUVAL}
\end{figure*}

\subsection{Local Training(Discriminate First)}

Pretraining a single model with audio and image modalities poses a unique challenge, as each modality reflects a distinct data pattern. Instead of using embeddings to convert the image and audio modalities \cite{girdhar2022omnivore}, we convert the audio to a 2-dimensional mel-spectrogram \cite{saeed2021contrastive} to construct a common representation of image and audio modality. We then resize the imaging modality to the size of Mel-spectrograms. 
This enables us to use both CNN and ViT for SSL pertaining. Our pretraining method is unique compared to other SSL methods. We denote the image data as $I_v =\{i_{v}^{m}\}_{m=1}^{M}$ and the audio data as $I_a =\{i_{a}^{n}\}_{n=1}^{N}$. Given the total dataset $D = \{I_{v} \cup I_{a}\}$ as input, the goal is to learn a function $f(.)$ that can map the input to a global representation feature space. We perform such a mapping sequentially in a batch-wise manner, as shown in Figure \ref{fig:FSSUVAL} (a),  with contrastive loss. 
\begin{equation}
   L_{SimCLR} = u^{T}v^{+}/\tau - log\sum_{v \in \{v^{+},v^{-}\}} exp(u^{T}v/\tau)  
\end{equation}
The above equation represents the NT-Xent (Normalized Temperature-scaled Cross Entropy) loss function as proposed in \cite{chen2020simple}. The input $u^{T}$, $v^{+}$, and $v^{-}$ are $l_{2}$-normalized. $\tau$ is a temperature coefficient. Note that the audio and image datasets are unpaired; therefore, each batch of data fed to the model can contain only images or audio but not both simultaneously. Such an approach enables the model to solve one SSL task (image features matching) in one batch and another SSL task (audio features matching) in another batch. By using such a training strategy, we aim to maximize the learning universality \cite{wang2024explicitly}: \textit{The learned representations for the audio and image should be distinct from each other, while the pretrained model should perform better or at least on par with the modality-specific models on their corresponding downstream tasks.} 

\subsection{Federated Learning}

We follow the approach of \cite{rehman2023dawa, kairouz2021advances} to train the federated version of the proposed SSL approach (see. Figure \ref{fig:FSSUVAL}). In Particular, we compose a dataset $D$ into $K$ partitions $\{d_{k}\}_{k=1}^{K}$, where each $k^{th}$ partition represents a decentralized client with $S_{a}=\{s_{j}\}_{j=1}^{J}$ audio samples and $S_{I}=\{s_{n}\}_{n=1}^{N}$ image samples, $J\neq N$. Note that this formulation represents an ideal scenario in which each client possesses both audio and image samples. However, they are still heterogeneous in terms of the uneven distribution of image and audio samples on the clients. In Section \ref{subsec:ch-of-FL}, we provide an in-depth investigation of the case where some of the clients do not possess one of the modalities, also shown in Figure \ref{fig:FSSUVAL}(b).

During each communication round $r$ of FL, the server randomly selects $Q=\{d_{q}\}_{q=1}^{Q}$ clients to participate in decentralized training and initialize the local models with the global model weights $w_{g}^{r}$. Each of the selected decentralized clients learns the global feature representations by training with SSL on its local datasets $d_{k}$ for a certain number of $E$ epochs before transmitting their local model to $w_{q}^{r}$ to the server. 

 \begin{equation}
    w_{q}^{r} = f_{SSL}(d_{k}, w_{g}^{r}, E)
 \end{equation}

 The server then receives the local models $\{w_{q}^{r}\}_{q=1}^{Q}$ from all selected clients and aggregates them based on a weighting factor $\beta(\cdot)$ to generate a new global model $w_{g}^{r+1}$ as follows:

\begin{equation}
    w^{r+1}_{g} = \sum_{q=1}^{Q}\beta_{q}w_{q}^{r}.
    \label{eq:1}
\end{equation}
This process is repeated until model convergence. We use FedAvg \cite{mcmahan2017communication} to aggregate the local models at the server, i.e., $\beta_{q} = \frac{d_q}{\sum_{q=1}^Q d_q}$.

\subsection{Learning Universality in FSSL} 
\begin{figure*}[htb]
    \centering
       \begin{subfigure}{0.25\linewidth}
         \includegraphics[width=\linewidth]{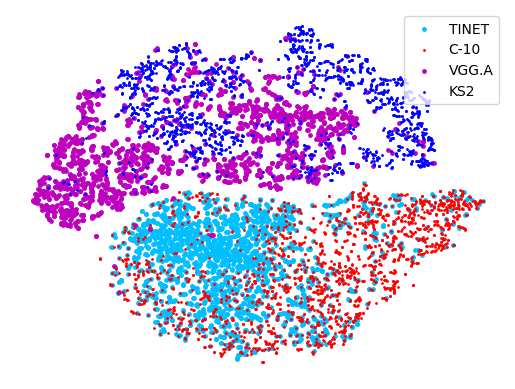}
    \caption{Round 10}
    \end{subfigure}%
    \begin{subfigure}{0.25\linewidth}
         \includegraphics[width=\linewidth]{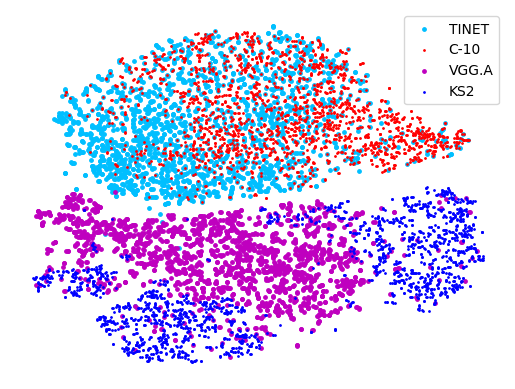}
    \caption{Round 40}
    \end{subfigure}%
     \begin{subfigure}{0.25\linewidth}
         \includegraphics[width=\linewidth]{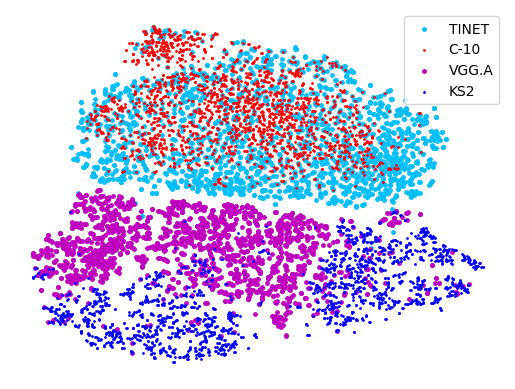}
    \caption{Round 70}
    \end{subfigure}%
     \begin{subfigure}{0.25\linewidth}
         \includegraphics[width=\linewidth]{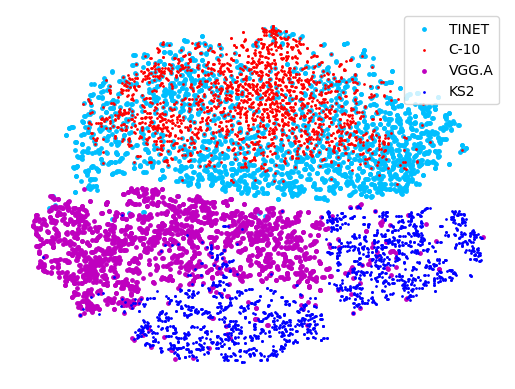}
    \caption{Round 100}
    \end{subfigure}


     \begin{subfigure}{0.25\linewidth}
         \includegraphics[width=\linewidth]{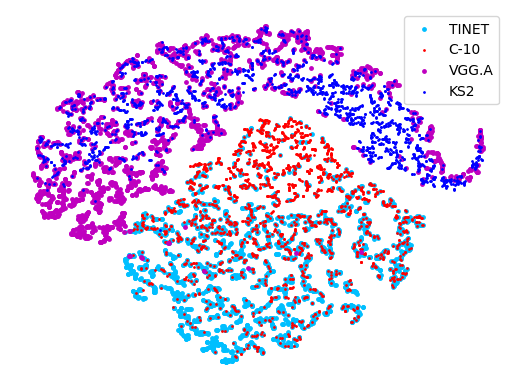}
    \caption{Round 10}
    \end{subfigure}%
    \begin{subfigure}{0.25\linewidth}
         \includegraphics[width=\linewidth]{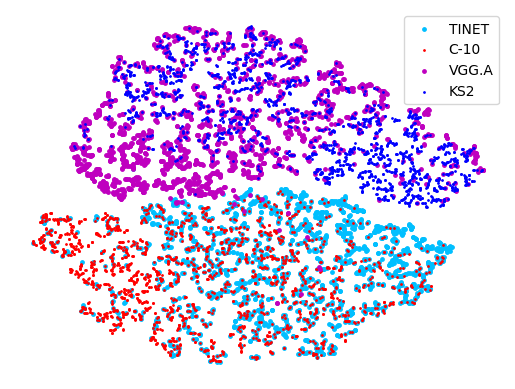}
    \caption{Round 40}
    \end{subfigure}%
     \begin{subfigure}{0.25\linewidth}
         \includegraphics[width=\linewidth]{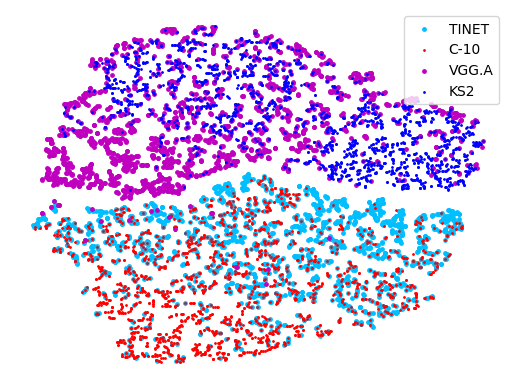}
    \caption{Round 70}
    \end{subfigure}%
     \begin{subfigure}{0.25\linewidth}
         \includegraphics[width=\linewidth]{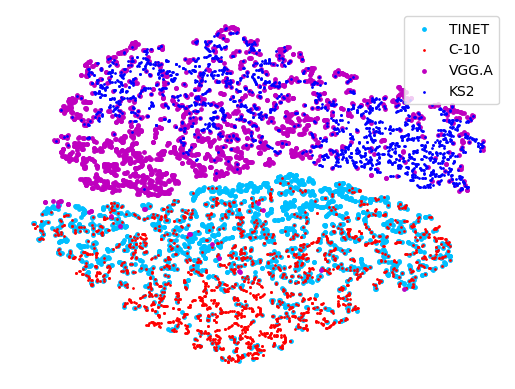}
    \caption{Round 100}
    \end{subfigure}

\caption{TSNE plot of the features of Tiny ImageNet (TINET), CIFAR-10 (C-10),  VGG Audio (VGG.A), and Speech command V2 (KS2) datasets at different rounds of FSSUAVL pertaining. (Top-Row) CNN (Bottom-Row) ViT. }
    \label{fig:FSSAVL_rounds_TSNE}
\end{figure*}

According to \cite{wang2024explicitly}, the SSL-pretrained model learns universal representations for the samples of each task, i.e., simultaneously maximizing the similarity between the augmented views of similar samples and pushing a part dissimilar samples (\textbf{discriminability}). This further entails that these SSL-pretrained models should be able to classify various fractions of the data (e.g., separate different audio and image datasets). We consider how well the global model in FSSL can separate the seen and unseen audio and image samples to assess its learning universality. Figure \ref{fig:FSSAVL_rounds_TSNE} shows the TSNE plot of the resultant global model in various rounds of FSSL training for CNN and ViT. The clustering of various audio and image data fractions can be readily observed as the FSSL training progresses, which shows that the FSSL-pretrained model can cluster both intra-modality and inter-modality data. This property is indeed helpful for edge devices that contain multimodal data but cannot accommodate multiple models due to constrained computational resources.

\subsection{Implementation Details}


\noindent\textbf{Encoder} We use ResNet18 \cite{he2016deep, tamkin2021viewmaker} (referred to as R-18 afterward) and ViT \cite{dosovitskiy2020image} as encoders for the image and audio modalities. We follow the methodology of \cite{rehman2023dawa} for FL pretraining. For ViT, we attach an MLP head to the output of the transformer layer to map them to a $128$ dimensional vector following the setup of SSAST \cite{gong2022ssast}.

\noindent\textbf{Encoding Modalities:}
For visual modality, we use the image augmentations proposed in \cite{tamkin2021viewmaker} to construct the augmented views of image samples during FL pretraining. We resize the image to $128\times128$ dimension before feeding it to the encoder. For preprocessing the audio modality, we follow the approach of \cite{rehman2024exploring} to convert a randomly chosen 2-second audio sampled at 16 kHz into spectrograms using 128-mel spectrogram bins. We deliberately resized the time dimension of the spectrogram to $128$ to match it, in resolution, with the imaging modality.  

\noindent\textbf{Pretraining Datasets:}
We pretrain the proposed method with TinyImageNet (referred to as TINET afterward)\cite{chrabaszcz2017downsampled} and VGG sound \cite{chen2020vggsound}  (referred to as VGG.A afterward) datasets. The TINET dataset consists of 100 classes with each class containing 500 images. VGG.A is a large-scale audio dataset extracted from YouTube videos with more than 200K audio clips, each 10 seconds long. The audio is sampled at 16KHz. There are 309 classes in this dataset. One of the reasons for choosing VGG.A dataset instead of AudioSet \cite{gemmeke2017audio} (as commonly used in centralized setups) is due to the former being well curated and having close correspondence between the sounds emitted by objects and the labels. This helps us to simulate a highly Non-IID audio dataset.

We generate the Non-IID version of VGG.A and TINET using Dirchilet coefficient $\alpha$, where a lower value indicates higher heterogeneity \cite{lubana2022orchestra, rehman2023dawa}. This results in randomly partitioned data sets of 100 shards for \textit{cross-device} settings that mimic 100 disjoint decentralized clients in FL. It should be noted that we generate the 100 shards for VGG.A and TINET separately. We combined them to get 100 clients, each containing heterogeneous unpaired audio and image data samples.  Note that this further induces two types of data heterogeneities: 

(1) The distribution of audio and image samples on each client varies significantly. (2) This results in model learning either more image-based features or more audio-based features.

\noindent\textbf{Downstream Task Dataset:} For visual downstream task evaluation we use TinyImageNet\cite{chrabaszcz2017downsampled} and CIFAR10 \cite{krizhevsky2009learning} datasets. For audio downstream tasks, we use VGG Audio\cite{chen2020vggsound}, EpicKitchen\cite{huh2023epic}, and SpeechCommand \cite{speechcommandsv2} datasets; see Table \ref{tab:datasets} for summary.

 \begin{table}[!htb]
     \centering
     \resizebox{\linewidth}{!}{
     \begin{tabular}{llcccc}
     \hline
         Dataset & Symbol &Task & \#Classes & \#Train& \#Val  \\
         \hline
         TinyImageNet\cite{chrabaszcz2017downsampled} & TINET&  Image cls. & 200 &100K &10K \\
         CIFAR-10 \cite{krizhevsky2009learning} &C-10 & Image cls. & 10 & 50K & 10K \\
         \hline
         VGG Sound \cite{chen2020vggsound} &VGG.A & Action cls. & 309 & 183.730K & 15.446K \\
         Epic Sound \cite{huh2023epic} & Epic.K&  Action cls. & 44 & 60.056K & 8.036K \\
         Speech Command-V2 \cite{speechcommandsv2} & KS2& Speech cls. & 35 & 84.848K & 4.890K \\
         \hline
     \end{tabular}
     }
     \caption{Transfer datasets to evaluate the FSSUAVL on image and audio modalities. The table shows the dataset, the corresponding symbol used in this paper, their task type, the number of classes (\#classes), training samples (\#Train), and validation samples (\#val) for each dataset. }
     \label{tab:datasets}
 \end{table}

 
\noindent\textbf{FL Pretraining.} FL pretraining lasts for 100 rounds. In each round, we select 10\% of the clients to participate in FL pretraining before aggregating the models at the server. Each local model, when selected by the server, trains itself for 5 local epochs. We keep the learning rate unchanged throughout the FL pretraining following \cite{rehman2023dawa}. Local pretraining of both R-18 and ViT is performed with a batch size of 256, an initial learning rate of 0.03 which decays according to a cosine schedule, weight-decay of $1\times 10^{-4}$. The temperature coefficient for SimCLR loss is set to $\tau=0.5$. 
For ViT, the size of the input patch was set to $16\times16$ resulting in 64 patches. The number of heads in the ViT was set to 8 and the number of transformer layers was set to 18. \\

\noindent\textbf{Evlauation:} Unless otherwise specified, we use KNN Monitor \cite{wu2018unsupervised}, with $t=0.1$ and $k=200$, for most of our evluations. We also use linear probes (L.P) and end-to-end fine-tuning (F.T) when comparing the performance of the models.
For linear-probe(L.P) and end-to-end fine-tuning (F.T) with R-18, we replace the last linear layer of the pretrained model with a task-specific linear layer. We train the model for 60 epochs with a learning rate of $3\times10^{-2}$ that is linearly decayed by a factor of 0.1 after the 60\% and 80\% epochs. We use a similar procedure for ViT, except that we take the mean of the output features produced by the ViT model following SSAST \cite{gong2022ssast}. Note that during the fine-tuning, we keep the image and spectrogram dimensions to $128\times128$.

%% file: sec/4_experiment.tex
\section{Experiments}

We compare FSSUAVL performance against modality-specific models that are also pretrained with FSSL. Here, we call the modality-specific image and audio models FVSSL and FASSL, respectively.

\begin{table*}
\resizebox{\linewidth}{!}{
\begin{tabular}{ll |ccc| ccc | ccc}   
\hline
Method & Arch.   & \multicolumn{3}{c}{KNN Classification} & \multicolumn{3}{c}{L.P} & \multicolumn{3}{c}{F.T} \\
\cline{3-11}
&  &  TINET & C-10 & Avg.  & TINET & C-10 & Avg.& TINET & C-10 & Avg.\\
\hline
FVSSL\cite{chen2020simple}	& R-18			&  12.27 	& 45.65 & 28.96	 & 31.61	&75.04	&53.33 & 56.85&  75.96 &  66.41 \\
\rowcolor{lightgray}FSSUAVL	& R-18	 	& \textbf{12.50}	& \textbf{46.47}	& \textbf{29.49}	& \textbf{32.41}	& \textbf{75.16} &\textbf{53.79}& \textbf{59.49}& \textbf{86.7} & \textbf{73.10}  \\
\hline

\hline

FVSSL\cite{chen2020simple}	& ViT			& \textbf{4.17}	& \textbf{31.4}	& \textbf{17.79} & 6.57 & \textbf{40.18} & \textbf{23.38}  & 32.25 &  69.35 & 50.8 	\\
\rowcolor{lightgray}FSSUAVL	& ViT			& 3.86	& 30.19	& 17.03 & \textbf{6.77 }& 39.18 & 22.98 & \textbf{32.89} & \textbf{69.80} & \textbf{51.35} 	\\
\hline

\hline
\end{tabular}
}
\caption{Comparison of FSSUAVL with expert models in cross-device FL settings on vision downstream tasks. Arch. represents architecture, and Avg. represents average accuracy. L.P represents linear fine-tuning, and F.T represents full-network fine-tuning. }
\label{tab:vision-only}
\end{table*}

\begin{table*}[ht]
    \centering
    \resizebox{\linewidth}{!}{
    \begin{tabular}{lll ccc |cccc |cccc}
    \hline
    Method & Arch.   & \multicolumn{3}{c}{KNN Classification} & \multicolumn{4}{c}{L.P} & \multicolumn{4}{c}{F.T}\\
    \hline
     &  &Epic.K & VGG.A & KS2 & Avg. &  Epic.K & VGG.A & KS2 & Avg.&  Epic.K & VGG.A & KS2 & Avg. \\
    \cline{4-10}
    \hline
    FASSL\cite{rehman2022federated} & R-18 & 29.46	& 10.43	& 14.36	& 18.08& \textbf{38.59} &  28.15 & 67.99 & 44.91 & \textbf{46.37} & \textbf{49.19} & 95.93 & \textbf{63.83} \\ 
    
    \rowcolor{lightgray}FSSUAVL	&R-18	& \textbf{30.34}	& \textbf{9.98}	& \textbf{20.69}	& \textbf{20.34}	& 35.89	& \textbf{30.62}	& \textbf{74.93} & \textbf{47.15} & 44.56	&48.67	& \textbf{96.00}	&63.08\\
     \cline{3-10}
    \hline
    \hline 
    FASSL\cite{rehman2022federated} & ViT & 20.09&  3.4& 11& 11.50 & 23.69  & 7.19 & 17.51 & 16.13 & \textbf{41.83} & \textbf{39.63} & \textbf{86.66} & \textbf{56.04} \\
    
    \rowcolor{lightgray}FSSUAVL&	ViT		&\textbf{21.58} &  \textbf{4.26} & \textbf{14.06} & \textbf{13.30}	&\textbf{24.45}	& \textbf{8.72}	& \textbf{22.35} & \textbf{18.52}  &41.31	&36.04	&85.17	&54.17\\

         \hline
    \end{tabular}}
    \caption{Comparison of FSSUAVL with expert models in cross-device FL settings on Audio downstream tasks. Arch. represents architecture, and Avg. represents average accuracy. L.P represents linear fine-tuning, and F.T represents full-network fine-tuning. }
    \label{tab:audio-only}
\end{table*}

\subsection{Unimodal Evaluation}

\noindent\textbf{Vision:} 
As shown in Table \ref{tab:vision-only}, FSSUAVL with R-18 in the 3 evaluation tests performs much better than FVSSL. In particular, FSSUVAL obtained an average performance improvement of 0.53\%, 0.46\%, and 6.69\% in KNN classification, L.P, and F.T, respectively, compared to FVSSL. On the other hand, FSSUAVL with VIT shows competitive performance against FVSSL in KNN classification and L.P and improved performance in F.T. These results suggest that the use of combined audio and image modalities during FSSL pretraining can improve the performance of image modalities.

\noindent\textbf{Audio:} 
Table \ref{tab:audio-only} shows the performance comparison of FSSUAVL against FASSL. One can see that FSSUVAL with R-18 and VIT shows performance improvement of 2.26\% and 1.8\%, respectively, against FASSL in KNN classification. Similarly, in L.P evaluation, FSSUAVL shows an average improvement of 2.24\% and 2.39\% for R-18 and VIT, respectively, against FASSL. In the case of F.T, we found that FSSUVAL shows competitive performance against FASSL with R-18 and a performance degradation of 1.87\%  with VIT. We conjecture that such performance improvement in audio-based downstream tasks is due to the FSSUAVL-pretrained model leveraging the learned image knowledge during local SSL training on audio data.

\begin{figure}[htb]
    \centering
     \begin{subfigure}{0.33\linewidth}
         \includegraphics[width=\linewidth]{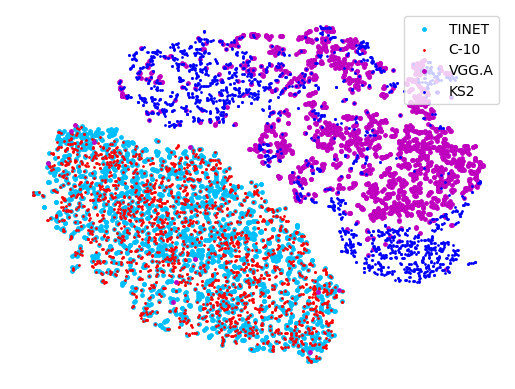}
    \caption{}
    \end{subfigure}%
       \begin{subfigure}{0.33\linewidth}
         \includegraphics[width=\linewidth]{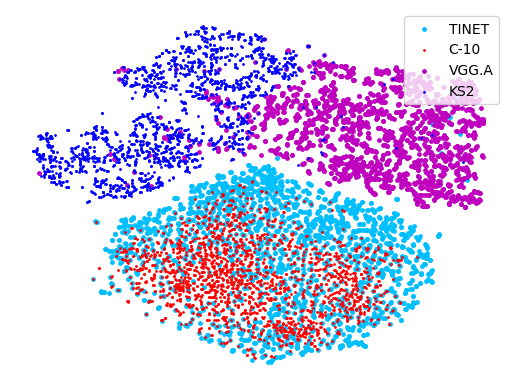}
    \caption{}
    \end{subfigure}%
    \begin{subfigure}{0.33\linewidth}
         \includegraphics[width=\linewidth]{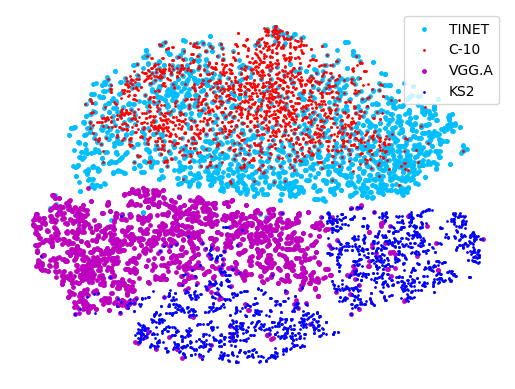}
    \caption{}
    \end{subfigure}
     \begin{subfigure}{0.33\linewidth}
         \includegraphics[width=\linewidth]{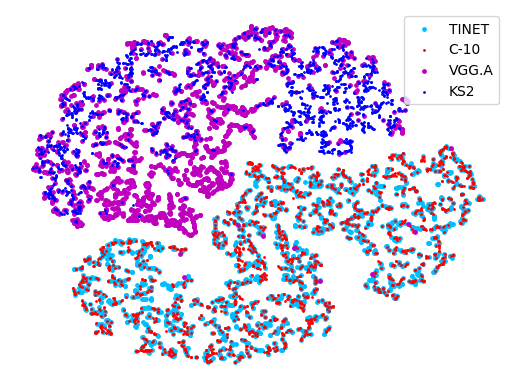}
    \caption{}
    \end{subfigure}%
     \begin{subfigure}{0.33\linewidth}
         \includegraphics[width=\linewidth]{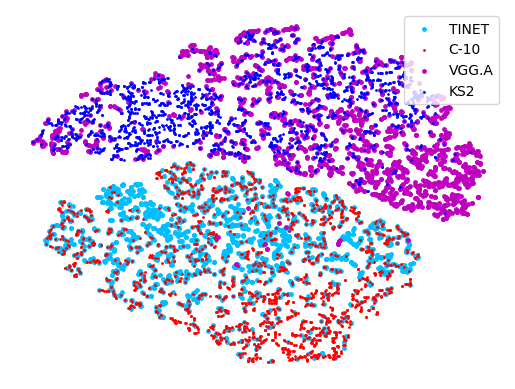}
    \caption{}
    \end{subfigure}%
     \begin{subfigure}{0.33\linewidth}
         \includegraphics[width=\linewidth]{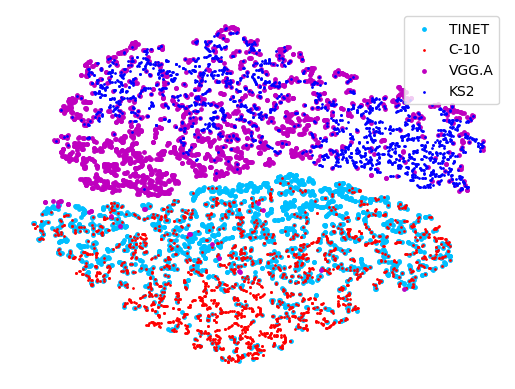}
    \caption{}
    \end{subfigure}

\caption{TSNE plot of the features of TINET, C-10, VGG.A and KS2 datasets after FSSL pertaining. (Top-Row) R-18 (Bottom-Row) ViT.  (a\& d)FASSL, (b \& e) FVSSL, and (c\& f) FSSUAVL pretraining  }
    \label{fig:FSSAVL_TSNE}
\end{figure}

\noindent\textbf{Learning Universality:} The results in Table \ref{tab:vision-only} and Table \ref{tab:audio-only} can be explained by the TSNE plots of FASSL, FVSSL, and FSSUAVL and observing how how well these methods separate different modalities in large-scale (TINET and VGG.A) and small-scale datasets (C-10 and KS2). 
As shown in Figure \ref{fig:FSSAVL_TSNE}, we found that pretraining R-18 and ViT with SSL using either image, audio, or both performed better in separating images and audio modality features, even when the audio-pretrained model has not seen any image samples and image pretrained model has not seen any audio samples. 
However, we found that the out-of-domain features for FASSL (See Figure \ref{fig:FSSAVL_TSNE} (a \& d)) with R-18 and ViTs are more spread out in the features space rather than clustered together, indicating that these uni-model SSL techniques struggle to find the similarity between different out-of-domain data samples.   On the other hand, we found that FVSSL and FSSUAVL, as shown in Figure \ref{fig:FSSAVL_TSNE} cluster each dataset's samples more efficiently with both R-18 and ViT. 
For unpaired datasets, separating various inter-modality and intra-modality features with a single model is cost-effective. 

\subsection{Performance on Audio-Visual Recognition Task}
One key advantage of FSSUAVL over modality-specific models is its ability to generalize across both image and audio modalities. This generalization occurs naturally since the same model is employed for both modalities. Notably, FSSUAVL is trained without paired audio-visual data or multimodal feature-matching loss \cite{girdhar2023imagebind}. We assess the performance of both modality-specific models and FSSUAVL on out-of-domain image and audio data representing the same class or scene. 

For this purpose, we use the ADVANCE \cite{hu2020cross}, and a combination of MNIST \cite{lecun2010mnist} and Free Spoken Digit (FSSD)  \cite{Spoken_Digit} datasets (MNIST+FSSD). The ADVANCE dataset consists of audio-visual paired data of geotagged aerial scenes. There are 13 classes in this dataset with a total of 5075 audio-image pairs. We use 70\% of the data for training the model and 30\% data for evaluating the model. To create an image audio pair from the combination of MNIST and FSSD datasets, we pair each audio of FSSD with the corresponding image in the MNIST dataset. As FSSD only has 3000 audio samples, it resulted in 2700 audio-image pairs for training and 300 audio-image pairs for testing corresponding to the digits from 0 to 9. For recognition, we sequentially fed the image and corresponding audio of the scene to the model. The category of the class is determined by taking the mean of the logits output by the model for the image and its corresponding audio modality. It is worth noting that the ADVANCE and MNIST+FSSD datasets represent completely different tasks. The former deals with the recognition of aerial imagery supplemented by corresponding audio event information while the latter represents recognizing the digits given both image and speech information.   

We report the transfer learning performance of FVSSL, FASSL, and FSSUAVL on these datasets in Table \ref{tab:ADVANCE_reslts}. The results demonstrate that FSSUAVL effectively leverages both image and audio modalities, outperforming single-modality approaches in most scenarios. With R-18, FSSUAVL significantly surpasses both FVSSL and FASSL across the ADVANCE and MNIST+FSSD datasets. For ViT, we observe similar advantages, with FSSUAVL maintaining competitive performance against FVSSL on the ADVANCE dataset while excelling on MNIST+FSSD. These findings further validate our unified multimodal approach, which harnesses complementary information from diverse modalities to enhance prediction reliability.

\begin{table}[t]
    \centering
    \resizebox{\linewidth}{!}{
    \begin{tabular}{c|ccc|ccc}
    \hline
          Arch. & \multicolumn{3}{c}{ADVANCE \cite{hu2020cross}} & \multicolumn{3}{c}{MNIST\cite{lecun2010mnist} + FSSD\cite{Spoken_Digit}}\\
          
          \cline{2-7}
         &  FVSSL \cite{chen2020simple} & FASSL \cite{rehman2022federated} & FSSUAVL & FVSSL \cite{chen2020simple} & FASSL \cite{rehman2022federated} & FSSUAVL \\
         \hline
        R-18 & 67.81 & 42.81 & \textbf{72.19} & 97.66 & 89.84 & \textbf{ 98.44}   \\
    \hline
    \cline{2-7}
    ViT & \textbf{43.75} & 31.89 & 42.19 & 39.84 & 39.06  &\textbf{ 53.13} \\
    \hline
    \end{tabular}
    }
    \caption{Transfer learning performance comparison of FVSSL, FASSL, and FSSUAVL on ADVANCE \cite{hu2020cross} and MNIST\cite{lecun2010mnist}+FSSD\cite{Spoken_Digit} datasets.}
    \label{tab:ADVANCE_reslts}
\end{table}

\subsection{Further Analysis on Audio-Visual Dataset}

We conducted additional evaluations on the ADVANCE \cite{hu2020cross} and MNIST \cite{lecun2010mnist} + FSSD \cite{Spoken_Digit} datasets to rigorously assess the effectiveness of our proposed method. In particular, we measure the performance of FVSSL, FASSL, the combination of FVSSL and FASSL (FVSSL +FASSL), and FSSUAVL when subjected to single-modality or paired-modality data, without further fine-tuning. We used KNN Monitor \cite{wu2018unsupervised} to quantify classification accuracy. The complete results are presented in Table \ref{tab:ADVANCE_AV}. For evaluations involving paired multimodal data (Audio+Visual), we created unified representations by concatenating the extracted image and audio features from each sample.

As shown in Table \ref{tab:ADVANCE_AV}, except for ViT in the ADVANCE dataset, FSSUVAL consistently outperformed the combination of FVSSL and FASSL (FVSSL+FASSL). 
This comparison is particularly significant considering that FASSL+FVSSL requires twice as many model parameters since it processes image and audio modalities through separate pretrained models\cite{feng2023fedmultimodal, girdhar2023imagebind}. In contrast, FSSUAVL achieves superior multimodal feature extraction using a single unified model, substantially reducing the computational and training costs associated with multiple expert models. The computational cost savings from our approach become more magnified when the number of modalities increases, making the deployment of combining multiple expert models of various sizes on edge devices infeasible \cite{girdhar2023imagebind, srivastava2024omnivec}.   
When evaluating single-modality data, FSSUAVL shows competitive performance compared to FVSSL and FASSL, demonstrating its versatility in audio and image modalities.

 \begin{table*}[!htb]
\centering
\resizebox{\linewidth}{!}{
\begin{tabular}{ccc|ccc|ccc|ccc|ccc}
\hline
         & & & \multicolumn{6}{c}{ADVANCE \cite{hu2020cross}} & \multicolumn{6}{|c}{MNIST \cite{lecun2010mnist}+FSSD \cite{Spoken_Digit}} \\
         \hline
          & & & \multicolumn{3}{c}{R-18} & \multicolumn{3}{|c}{ViT} & \multicolumn{3}{c}{R-18} & \multicolumn{3}{c}{ViT}\\
          \cline{4-15}
         Method &  P.T.D & Parameters (R-18/ViT) & A+V & V & A & A+V & V & A & A+V & V & A & A+V & V & A  \\
         \cline{4-15}
         \hline

        FVSSL & TINET & 12M/15M &  \xmark & \textbf{53.74} & \xmark &  \xmark & 40.83 & \xmark & 
                     \xmark & 68.36 & \xmark & \xmark & \textbf{26.17}&  \xmark \\
        FASSL & VGG.A &12M/15M & \xmark & \xmark & \textbf{27.51} & \xmark & \xmark &  24.80 & 
                    \xmark & \xmark &  48.05 & \xmark & \xmark & 28.52\\
                    \hline
        FASSL + FVSSL & TINET, VGG.A & 24M/30M & 51.15 &  \xmark & \xmark & \textbf{42.39} & \xmark &  \xmark  & 
                            67.58 &  \xmark & \xmark & 37.50 & \xmark & \xmark\\
         
         \rowcolor{lightgray}FSSUAVL & VGG.A + TINET & \textbf{12M/15M }& \textbf{52.17} & 52.92 &25.14 & 35.52 & 34.65 &\textbf{ 25.0} &
         \textbf{ 66.41} & 62.89 & \textbf{55.86} & \textbf{37.89} & 20.70 &\textbf{41.80}\\
         \hline

         \hline
    \end{tabular}
    }
    \caption{\% KNN classification performance of FVSSL, FASSL, and FSSUAVL with paired audio-visual (A+V), only visual (V), and only audio (A) modalities. P.T.D represents the pretraining dataset.  }
    \label{tab:ADVANCE_AV}
\end{table*}

\section{Ablation Studies}
\label{subsec:ch-of-FL}

\noindent\textbf{Charcteristics of FL.}
 We evaluated FSSUAVL in various FL setups since such models can be advantageous in these scenarios where the devices are computationally constrained and the data is highly heterogeneous. 
\subsection{Model Combination vs. FSSUAVL.}
Data heterogeneity in FL is known to severely affect the performance of the global model \cite{lubana2022orchestra, rehman2023dawa, mcmahan2017communication}. The presence of audio and visual modality on edge devices offers an additional layer of data heterogeneity, where some edge devices may contain both audio and image modalities, while other edge devices are unimodal (clients with only image or audio modalities, but not both). We evaluated how effectively FSSUAVL leverages the clients with both audio and image modality in the presence of unimodal clients.    

As a baseline, we consider the case where we restrict the client's model to learn only from either image or audio modality but not both at the same time. This restricts FSSUAVL to only model combinations. Table \ref{tab:FL_CASE} shows the performance for the simple model combination against FSSUVAL. It can be observed that FSSUAVL performs much better than the model combination by leveraging the clients containing both audio and image data. We found that the performance degradation for the model combination is more severe on the visual-based downstream tasks and semantic-audio downstream tasks, which explains the reasons for the requirement of additional resources on the clients \cite{sun2024towards} and at the server (such as hyper-networks)\cite{chen2024fedmbridge}. In contrast, FSSUAVL takes advantage of the availability of clients with unpaired image and audio data and uses the same model to process both audio and visual data, providing simplicity and effectiveness in processing unpaired multimodality data with acceptable performance. This further shows that a single model with SSL can be utilized to achieve gains in performance in scenarios with high data heterogeneity, and those gains could be further boosted by using auxiliary methods. We further evaluate this special case of FSSUAVL in the subsequent sections as it is closer to the practical scenarios.

\begin{table*}
\centering
\resizebox{1\linewidth}{!}{
\begin{tabular}{lccccccccc}
\hline
Method&	Arch. &	 Clients w/ Audio only &	 Clients w/ Image only &	Clients w/ Audio \& Image &	TINET & 	C-10	& Epic.K & VGG.A &  KS2\\
\hline

Model-Comb. &	R-18& 50 & 50 & 0 &	4.01&	31.42 &	29.25& 10.14	&	14.79\\

FSSUAVL&	R-18&		33 &	33 & 33 & 	\textbf{10.37}& 	\textbf{42.76}&	\textbf{31.46} &	\textbf{10.92} &	\textbf{25.23}\\
\hline
Model-Comb. &	ViT& 	50& 	50 & 	0&	2.71&	28.17 &	19.26&	3.61&	12.33\\
FSSUAVL &	ViT& 	33& 	33& 	33 & \textbf{3.61}	& \textbf{30.37} &	\textbf{21.73} &	\textbf{4.07} &	\textbf{12.27}\\

\hline
\end{tabular}
}
\caption{Performance of FSSUAVL against simple model combination }
\label{tab:FL_CASE}
\end{table*}

\subsection{Effect of Varying Audio and Visual Data on the Clients.}
\begin{figure}[htb]
    \centering
    \begin{subfigure}{0.5\linewidth}
    \centering
         \includegraphics[width=0.9\linewidth]{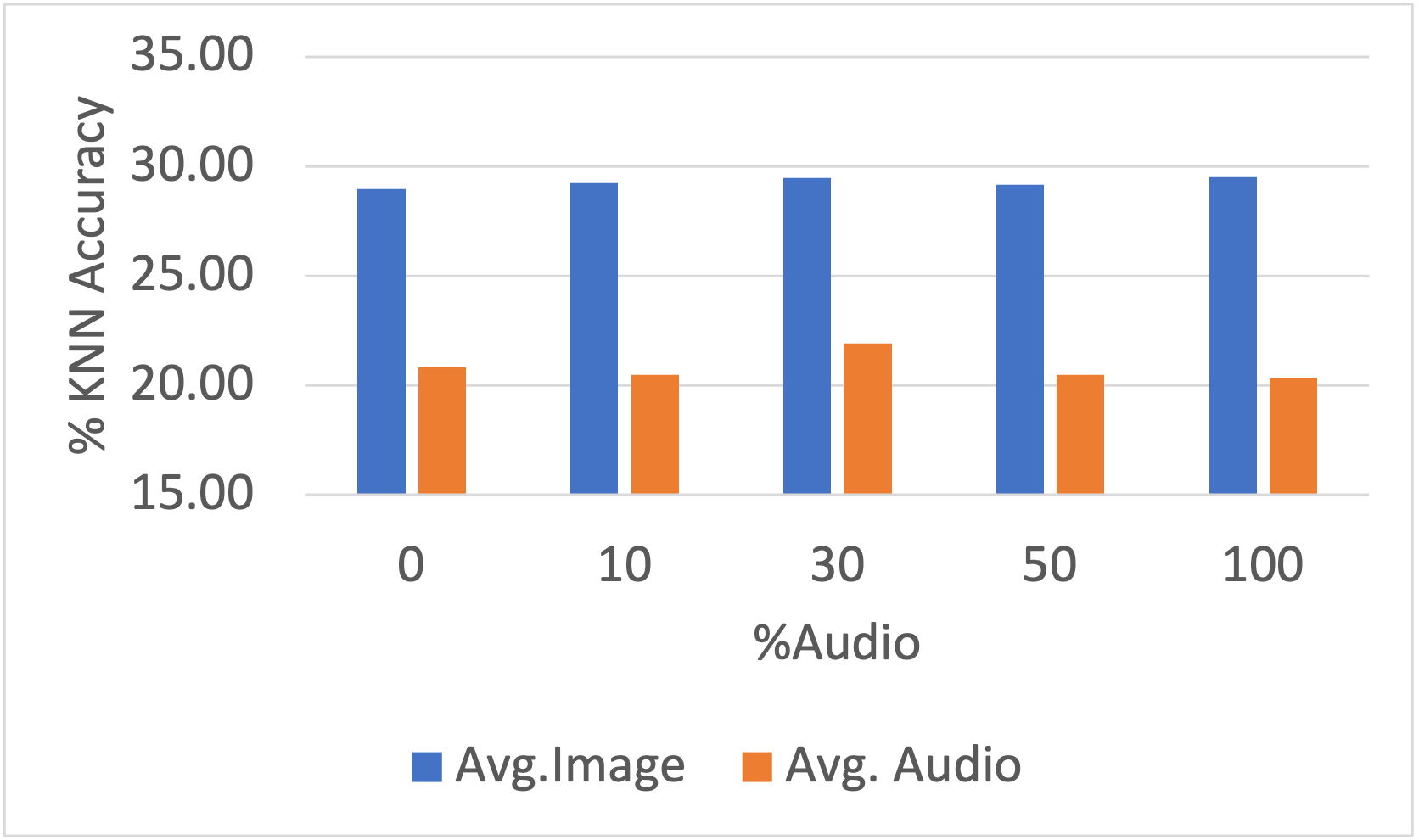}
    \caption{}
    \end{subfigure}%
    \begin{subfigure}{0.5\linewidth}
    \centering
         \includegraphics[width=0.9\linewidth]{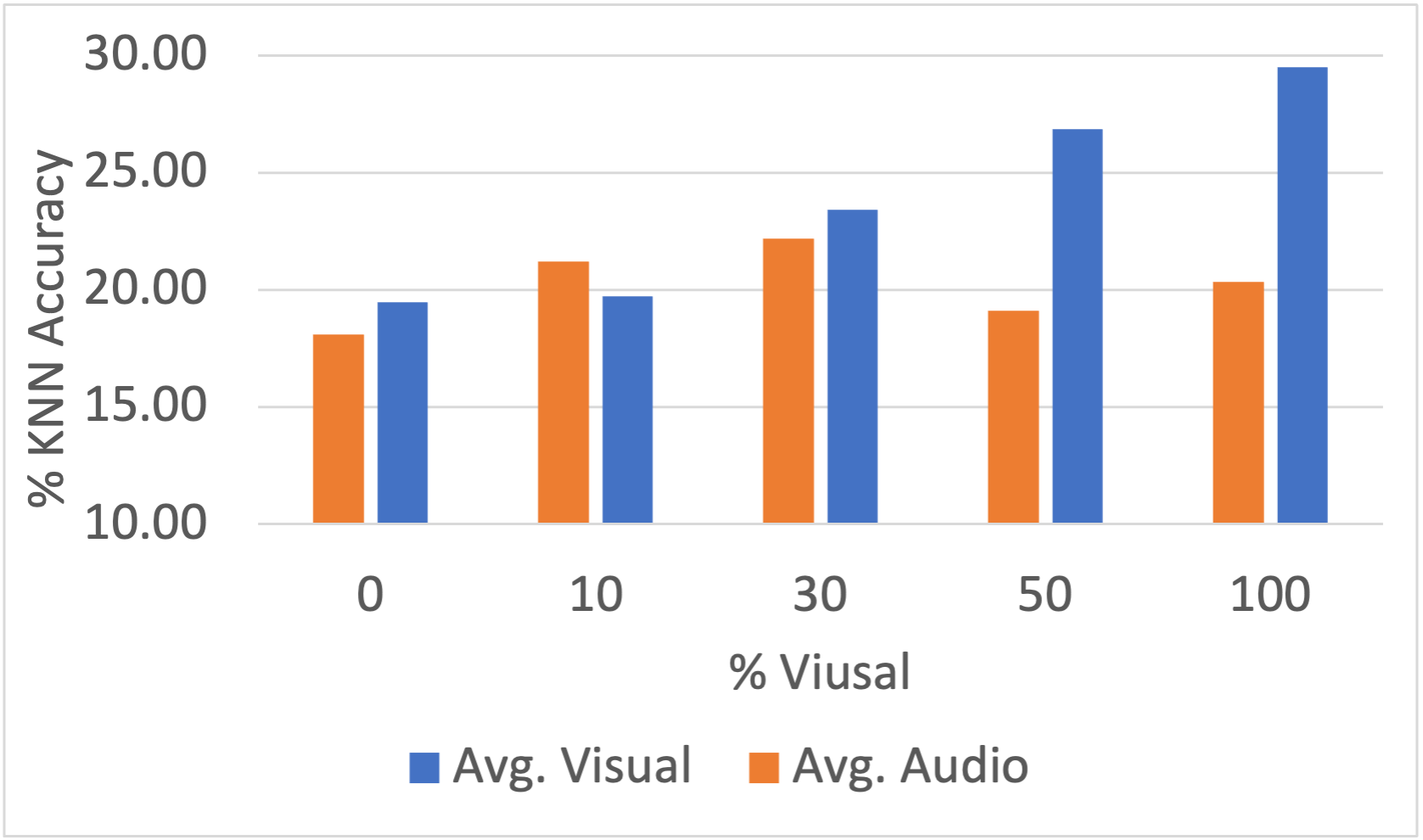}
    \caption{}
    \end{subfigure}
   \caption{Average performance of FSSUAVL with \% varying audio and visual data on the clients. (a) Varying \% of audio data while keeping the visual data at 100\% on the clients. (b) Varying the \% of visual data while keeping the audio data at 100\% on the clients.}
    \label{fig:FSSUAVL_change_av}
\end{figure}

To simulate the case where the clients contain varying percentages of audio and visual data, we started by keeping only image or audio data on all clients and then progressively adding audio or image data to a certain percentage of the clients. This represents the scenarios in which a client device may initially contain only one type of data and later acquire the data of other modalities.
Figure \ref{fig:FSSUAVL_change_av} (a) shows that the average optimal performance in the audio downstream tasks, with 100\% of visual-data-clients, is achieved when at least 30\% of the clients contain both audio and image data. Surprisingly, we found in Figure \ref{fig:FSSUAVL_change_av} (b) that a similar conclusion holds for the audio downstream task when 100\% of clients contain audio data and 30\% of clients hold both audio and image data.  
At the same time, Figure \ref{fig:FSSUAVL_change_av}(a) and (b) show that the average optimal performance on the visual downstream task is obtained when at least 50\% or 100\% of the clients contain both audio and image data. 
These results further show the effectiveness of FSSUAVL with vanilla FedAvg. 

\subsection{FSSUAVL with Barlow Twins.}
We further compare the performance of FSSUAVL with SimCLR and Barlo Twins \cite{zbontar2021barlow}. Barlow Twins is similar to SimCLR, with the exception that it does not require additional negative samples when computing the loss. We use similar settings as for SimCLR, including setting the temperature coefficient $\tau=0.5$.  For FSSL pretraining, we consider the case where only 33\% of the clients contain unpaired audio and image data, while the rest of the clients only contain image or audio data but not both. The FL pretraining lasts for 100 rounds, where in each round, 10\% of the local models are trained for 1 local epoch. We use the KNN monitor \cite{wu2018unsupervised} to report performance in Table \ref{tab:ssavl_other_SSL}.  One can see that the average performance of FSSUAVL with Barlow Twins and SimCLR on audio-based tasks is nearly similar, except for R-18 with vision-based downstream tasks, where we found that SimCLR performs much better than Barlow Twins.

\begin{table}
\centering
\resizebox{\linewidth}{!}{
\begin{tabular}{l  c c c c c c c c c} 
\hline
Method & Arch. & SSL & TINET  & C-10 & Avg. & Epic.K & VGG.A & KS2 & Avg. \\
\hline
FSSUAVL & R-18 & Barlo Tiwns  & 4.06 & 28.74 & 16.4 & 35.84 & 18.06 & 26.48 &  26.79\\
FSSUAVL & R-18 & SimCLR  & 3.77 & 34.7 & 19.24& 34.7 & 16.7 & 27.51 & 26.30 \\
\hline
FSSUAVL & ViT & Barlo Twins  & 2.47 & 23.18 & 12.83  & 26.55 & 9.04 & 15.20 & 16.93 \\
FSSUAVL & ViT & SimcLR  & 2.45 & 22.82 & 12.64 & 27.15 & 9.46 & 15.61& 17.40 \\

\hline
\end{tabular}
}
\caption{\% KNN classification accuracy of FSSUAVL with SimCLR and Barlo Twins. Arch. represents architecture. Avg. represents the average score.}
\label{tab:ssavl_other_SSL}
\end{table}

\section{Comparison with SOTA methods}
We compare the transfer learning performance of FSSUAVL against state-of-the-art (SOTA) FL vision-based and audio-based methods in Table \ref{tab:comp_cent_V}. On the TINET and C-10 datasets, FSSUAVL with R-18 achieves superior performance, outperforming previous FL methods, with a notable accuracy of 75.16\% on C-10 compared to the best prior result of 71.58\% from ORCHESTRA \cite{lubana2022orchestra}. Similarly, on the Epic.K dataset, FSSUAVL with R-18 and ViT demonstrates competitive results against Rehman et al. \cite{rehman2024exploring} (35.89\% and 24.25\% vs. 38.59\% and 23.69\%, respectively), while significantly surpassing the same method on the VGG.A (30.62\% and 8.72\% vs. 28.15\% and 7.19\%) and KS2 datasets (74.93\% and 22.35\% vs. 67.99\% and 17.51\%).
It is worth noting that prior image-based FL (cross-device) methods, such as ORCHESTRA, FedEMA\cite{zhuang2022divergence}, L-DAWA\cite{rehman2023dawa}, FedU\cite{zhuang2021collaborative}, and FedAnchor\cite{qiu2024fedanchor}, typically train and evaluate on the same dataset (e.g., C-10), overlooking out-of-domain downstream task evaluation. In contrast, FSSUAVL, pretrained on TINET and VGG.A, provides a comprehensive evaluation of R-18 and ViT across both in-domain (e.g., TINET, VGG.A) and out-of-domain (e.g., C-10, Epic.K, KS2) image and audio downstream tasks, showcasing its robustness and versatility.

\begin{table}
    \centering
    \resizebox{\linewidth}{!}{
    \begin{tabular}{lcccc|ccc}
    \hline
    Method & Arch.   & PTD.  & TINET & C-10 & Epic.K & VGG.A & KS2  \\ \hline    
\hline

\hline
 ORCHESTRA \cite{lubana2022orchestra} & R-18	 & C-10 & \xmark & 71.58  & \xmark & \xmark & \xmark \\
    
    FedEMA \cite{zhuang2022divergence} & R-18   & C-10 &  \xmark & 64.19  &\xmark & \xmark & \xmark \\
    L-DAWA \cite{rehman2023dawa}& R-18  & C-10 & \xmark & 68.20 & \xmark  & \xmark & \xmark \\ 
    FedU \cite{zhuang2021collaborative} &R-18 & C-10 & \xmark & 68.52 & \xmark & \xmark & \xmark   \\  
    FedAnchor \cite{qiu2024fedanchor} & R-18    & C-10 & \xmark & 62.94 & \xmark & \xmark & \xmark \\
    Rehman et al.\cite{rehman2024exploring} &R-18 $\dag$ & VGG.A  & 9.70 & 44.39 & 38.59 & 28.15 & 67.99 \\
      Rehman et al.\cite{rehman2024exploring} &ViT $\dag$  & VGG.A & 1.52 & 24.20 & 23.69 & 7.19 & 17.51 \\
    \hline
    
    \rowcolor{lightgray}FSSUAVL &R-18 & TINET+ VGG.A 	&32.41	&\textbf{75.16} & 35. 89 & 30.62 & 74.93\\

    \rowcolor{lightgray}FSSUAVL &ViT	& TINET+ VGG.A & 6.77	&39.18  & 24.25 & 8.72 & 22.35  \\
    \hline
    \end{tabular}
    }
    \caption{Comparison of FSSUAVL with the visual and audio expert models FL (\textit{cross-device}) settings. $\dag$ represents the results that were reproduced. PT.D represents the pretraining dataset. }
    \label{tab:comp_cent_V}
    \end{table}

%% file: sec/6_conclusion.tex
\section{Conclusion and Future Work}
In this work, we proposed a novel approach for jointly training unpaired image and audio data using a single model within the FL framework. Our method, FSSUAVL, sequentially trains a single architecture using SSL for both audio and visual modalities that are distributed across clients. 
It consistently achieves superior performance in audio-based and image-based downstream tasks compared to modality-specific models. FSSUAVL demonstrates particular efficacy in FL environments where clients face significant communication and computational constraints. Notably, our experiments reveal that FSSUAVL maintains remarkable performance stability even under extreme Non-IID data distributions, including scenarios where certain clients completely lack one modality type. Future research will extend FSSUAVL to incorporate additional modalities beyond audio and images, including video, text, and hyperspectral imaging, further exploring its potential as a comprehensive multimodal learning framework in distributed settings.